\documentclass[9pt, sigconf]{acmart}
\AtBeginDocument{%
  }

\usepackage{hyperref}
\usepackage{multirow}
\usepackage{subfigure}
\usepackage{wrapfig}
\usepackage{xspace}
\usepackage{algorithm}
\usepackage{algpseudocode}
\usepackage{lipsum}
\usepackage{algorithm,algpseudocode,float}
\usepackage{subcaption}
\usepackage{graphicx}
\usepackage{makecell}
\usepackage{enumitem}
\usepackage[skip=3pt]{caption}

\usepackage{etoolbox}
\newcommand{\zerodisplayskips}{%
  \setlength{\abovedisplayskip}{3pt}%
  \setlength{\belowdisplayskip}{3pt}%
  \setlength{\abovedisplayshortskip}{3pt}%
  \setlength{\belowdisplayshortskip}{3pt}}
\appto{\normalsize}{\zerodisplayskips}
\appto{\small}{\zerodisplayskips}
\appto{\footnotesize}{\zerodisplayskips}

\setcopyright{none}

\renewcommand\footnotetextcopyrightpermission[1]{}

\begin{document}
\thispagestyle{plain}
\pagestyle{plain}

\newcommand{\name}{SpeakAssis\xspace}
\renewcommand{\algorithmicrequire}{\textbf{Input:}}
\renewcommand{\algorithmicensure}{\textbf{Output:}}
\title{Talk to Me, Not the Slides: A Real-Time Wearable Assistant for Improving Eye Contact in Presentations}

\author{Lingyu Du}
\affiliation{%
  \institution{Delft University of Technology}
    \city{Delft}
   \country{The Netherlands}
}
\email{lingyu.du@tudelft.nl}

\author{Xucong Zhang}
\affiliation{%
  \institution{Delft University of Technology}
    \city{Delft}
   \country{The Netherlands}
}
\email{xucong.zhang@tudelft.nl}

\author{Guohao Lan}
\affiliation{%
  \institution{Delft University of Technology}
  \city{Delft}
   \country{The Netherlands}
}
\email{g.lan@tudelft.nl}

\renewcommand{\shortauthors}{Anonymous Author(s)}

\begin{abstract}
Effective eye contact is a cornerstone of successful public speaking. It strengthens the speaker's credibility and fosters audience engagement. Yet, managing effective eye contact is a skill that demands extensive training and practice, often posing a significant challenge for novice speakers. In this paper, we present \name, the first real-time, in-situ wearable system designed to actively assist speakers in maintaining effective eye contact during live presentations. Leveraging a head-mounted eye tracker for gaze and scene view capture, \name continuously monitors and analyzes the speaker’s gaze distribution across audience and non-audience regions. When ineffective eye-contact patterns are detected, such as insufficient eye contact, or neglect of certain audience segments, \name provides timely, context-aware audio prompts via an earphone to guide the speaker's gaze behavior. We evaluate \name through a user study involving eight speakers and 24 audience members. Quantitative results show that \name increases speakers' eye-contact duration by 62.5\% on average and promotes a more balanced distribution of visual attention. Additionally, statistical analysis based on audience surveys reveals that improvements in speaker's eye-contact behavior significantly enhance the audience's perceived engagement and interactivity during presentations.

\end{abstract}



\begin{teaserfigure}
\centering
  \includegraphics[width=0.88\textwidth]{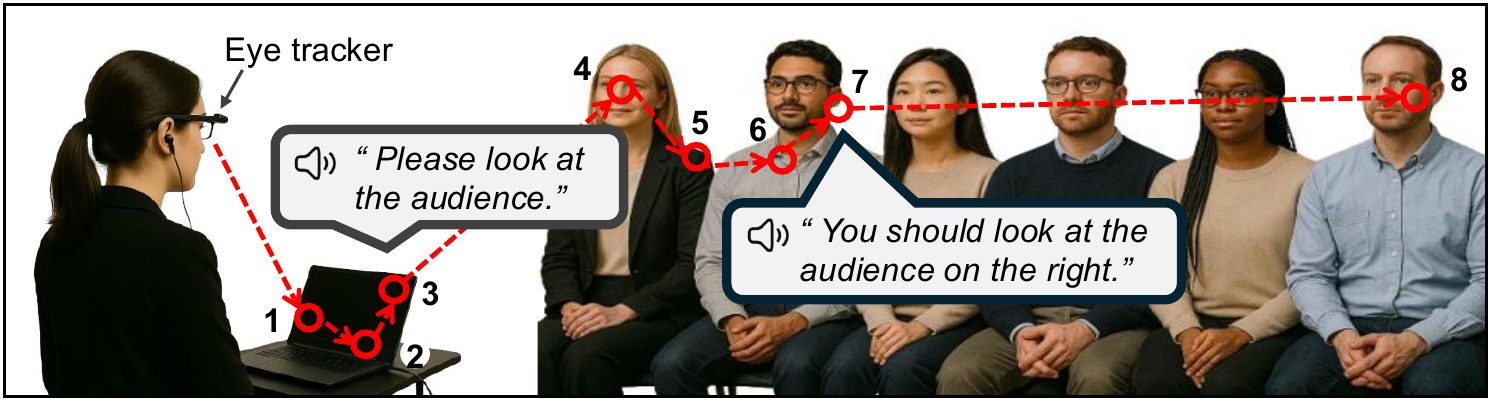}
  \caption{An illustration demonstrating how \name supports the speaker in maintaining effective eye contact in a presentation. \name continuously analyzes the speaker’s visual attention and provides real-time feedback to promote balanced engagement with the audience. The red dashed line indicates the speaker’s fixation trajectory. Initially, the speaker focuses on the laptop (fixations 1–3). \name detects this and provides an eye-contact prompt: ``\textit{Please look at the audience.}" The speaker then shifts attention to the audience but focuses only on the left side (fixations 4–7). In response, \name advises: ``\textit{You should also look at the audience on the right.}" The speaker then redirects her gaze to the right side (fixation 8).}
  \label{fig:teaser}
\end{teaserfigure}

\maketitle

\section{Introduction}

Eye contact is one of the most powerful and essential non-verbal cues in public speaking~\cite{gareis2006guidelines}. Establishing effective eye contact allows speakers to capture and hold the attention of the audience, build a sense of connection, and enhance overall engagement~\cite{barati2015impact}. Studies have shown that speakers who maintain meaningful eye contact are perceived as credible, projecting greater competence, honesty, and confidence~\cite{beebe1974eye}. By contrast, insufficient or poorly managed eye contact is often interpreted as a sign of anxiety~\cite{wortwein2015automatic, lee2019public, campbell2013public} or lack of self-confidence~\cite{nadiah2019students, nash2016if}, which can undermine the speaker’s authority and diminish the effectiveness of the speech~\cite{wortwein2015multimodal}. Given its importance, establishing effective eye contact is widely regarded as a core element of successful public speaking~\cite{barati2015impact} and is frequently emphasized in performance evaluations by professional public speaking organizations such as the Toastmasters~\cite{chollet2015exploring}. 

Establishing effective eye contact involves more than simply looking at the audience; skilled speakers maintain long eye contact, avoid over-fixation on one individual, and actively distribute their visual attention across all parts of the room to ensure everyone feels acknowledged~\cite{gareis2006guidelines, kilag2023use, cortina2015low, van2014first}. However, this skill is not innate~\cite{hart2017virtual}. Learning to manage gaze under the cognitive and emotional load of public speaking often requires deliberate training and repeated practice~\cite{wortwein2015multimodal}. Many novice speakers struggle to internalize this skill, especially in the dynamic setting of a live presentation.


A range of attention-tracking systems have been developed to provide feedback on speaker's gaze behavior~\cite{chollet2015exploring, chollet2016multimodal, coskun2021investigation, wortwein2015automatic}. For example, virtual environment-based training platforms~\cite{chollet2015exploring, chollet2016multimodal, wortwein2015automatic} simulate virtual audience members that offer simple and immediate feedback, such as having the virtual avatars nodding or smiling when the speaker makes eye contact. Additionally, wearable eye tracking-based solutions~\cite{coskun2021investigation, wearablesmarter} allow speakers to record their gaze behavior during a talk and review it offline to reflect on attention patterns for future improvement. While these solutions are effective for offline training, they are not designed to support speakers during live presentations. They lack real-time, in-situ feedback that would help speakers adjust their gaze behavior on the fly. As a result, speakers have to rely on prior training and internalized strategies that are insufficient for novice presenters under pressure.

To fill this gap, we present \name, the first real-time, in-situ wearable assistant designed to help speakers improve eye contact during live presentations. Leveraging the scene view and gaze signal captured by a head-mounted eye tracker, \name continuously monitors and analyzes the speaker's gaze distribution across all audience members and non-audience regions (e.g., laptop, ceiling, background). When ineffective eye-contact patterns are detected, such as prolonged focus on non-audience regions, over-fixation on a single audience, or neglecting audience members seated at the periphery, \name generates and delivers timely and context-aware eye-contact prompts to the speaker through a wireless earphone. Figure~\ref{fig:teaser} provides an example of how \name helps the speaker in a live presentation. When the system detects that the speaker is constantly focusing on the laptop without eye contact with audience, it issues a prompt \textit{``look at the audience"}; when the speaker consistently neglects audience seated on the right side of the room, a prompt such as \textit{``look audience on the right"} will be generated. This real-time guidance allows the speaker to make immediate adjustments and maintain effective eye contact throughout the presentation.  

Providing appropriate and timely guidance requires the system to accurately track the speaker’s gaze targets in real time. Specifically, the system must identify which audience member the speaker is looking at, referred to as the \textbf{\textit{target face}}, and monitor how the speaker's gaze is distributed over time. However, this is particularly challenging in live presentations, due to consistent head movements of the speaker, dynamic scene views, and variations in audience members' facial appearances caused by changing camera angles and occlusions. Additionally, audience members may move or appear intermittently in the scene view, further complicating reliable tracking across frames.

To overcome this challenge, we develop a novel identification method that leverages the relative spatial positions among audience members to identify the {\textit{target face}}. Specifically, we combine face identification with face tracking to efficiently recognize a reference audience member in the scene view, designated as the \textbf{\textit{anchor face}}. Once the \textit{anchor face} is established, the identity of the \textit{target face} is inferred based on its relative position to the \textit{anchor face}. This approach allows \name to robustly and efficiently track the speaker's attention across different audience members and non-audience regions, even under frequent camera shifts and variations in audience's facial appearances.

We implement \name using commercially available hardware: an off-the-shelf wearable eye tracker for scene and gaze captures, a laptop (typically used to display slides) for data processing, and a single earphone to deliver discreet audio prompts. A simple interface is provided to ensure accessibility and ease of use. In summary, our major contributions are three-fold:

\begin{itemize}[leftmargin=*, wide, labelwidth=!, labelindent=0pt]
\setlength\itemsep{0.3em}

    \item We propose \name, the first real-time, in-situ wearable system that provides live, actionable feedback to help speakers improve eye contact during public presentations. 

    \item We introduce a robust identification technique that identifies the speaker's gaze target by leveraging the relative spatial relationships within the audience layout. It enables continuous tracking of the speaker's gaze distribution across all the audience members and non-audience regions in the scene view. These capabilities are critical for the system to provide reliable eye-contact prompts under frequent head movements and dynamic scene changes.

    \item We conduct a comprehensive user study involving 28 participants (four speakers and 24 audience members) to evaluate \name in real presentation scenarios. Our quantitative analysis shows that \name increases speakers’ eye-contact duration by 62.5\% on average and promotes a more balanced distribution of visual attention. Moreover, statistical analysis, based on post-study audience surveys, confirms that improvements in speakers' eye contact behaviors significantly enhance the audience's perceived engagement and interactivity. 
    
\end{itemize}

\section{Related Work}


\subsection{Eye Tracking and Its Applications}


Eye tracking systems are broadly classified into two categories: remote eye tracking and near-eye eye tracking. Remote eye tracking uses a camera positioned at a distance from the user and can be implemented on various devices, including mobile phones~\cite{huynh2021imon}, laptops~\cite{Zhang_2015_CVPR}, webcams~\cite{Zhang2020ETHXGaze}, and tablets~\cite{huang2017tabletgaze}. However, remote systems often suffer from limited tracking accuracy and are highly sensitive to ambient lighting conditions. By contrast, near-eye eye tracking employs cameras placed close to the eyes, typically integrated into wearable head-mounted devices such as smart glasses~\cite{kassner2014pupil, tobiiProGlass, smi}. This setup provides significantly higher tracking accuracy and greater robustness to head movement and user mobility, making it ideal for use in dynamic and uncontrolled environments. In this work, we adopt a head-mounted eye tracker to capture accurate gaze data from the speaker and focus on designing real-time assistance to support effective eye contact during public speaking.

Advancements in eye-tracking technology have enabled a wide range of applications across diverse domains. In particular, eye tracking has been extensively used in cognitive sensing systems, owing to the strong correlation between eye movements and cognitive processes underlying visual perception. Notable applications include the early detection of autism spectrum disorder~\cite{guillon2014visual, shishido2019application}, mental workload estimation~\cite{pfleging2016model, yamada2018detecting}, and driver attention monitoring~\cite{tobiiDriverMonitor, bmwForbes}. In addition, gaze also serves as a fast and intuitive modality for interaction~\cite{Majaranta2014}, making eye tracking a valuable component in human-computer interaction (HCI). For instance, Gaze-Guided Narratives~\cite{GazeGuidedNarratives} assists users in locating reference objects, such as buildings in a city panorama, by guiding their gaze. G-VOILA~\cite{GVOILA} leverages large language models to support gaze-based information retrieval in everyday scenarios. CASES~\cite{CASES} analyzes users’ gaze patterns while reading to provide contextual assistance. Eye tracking has also been employed to study speakers’ gaze behavior during public speaking~\cite{van2014first, coskun2021investigation}, which offers insights into how gaze patterns affect audience engagement. Building on these effort, our work investigates gaze-based wearable assistance to support speakers in improving eye contact during live presentations.

\subsection{Speaking Assistance}

Our work is related to prior research aimed at assessing and improving speaking performance~\cite{chollet2018influence,siddiqui2023manifest, bartyzel2025exploring, palmas2021virtual, zhou2021virtual, PicturetheAudience, strangert2008makes, wortwein2015multimodal, chollet2015exploring, chollet2016multimodal, tanveer2015rhema, takac2019public, bachmann2023virtual, sulter2022speakapp, laske2022efficacy}. For instance, Strangert et al.\cite{strangert2008makes} analyzed acoustic prosodic features to evaluate speaker performance, showing that effective speakers often demonstrate fluent speech, an appropriate speech rate, and dynamic pitch variation. To support vocal delivery, Tanveer et al.~\cite{tanveer2015rhema} developed a feedback system using Google Glass that monitors speaker's volume and speaking rate to generate real-time verbal prompts such as ``louder'' and ``faster'' to guide speakers during presentations.

Other works have explored the use of virtual environments to train speakers in maintaining effective eye contact~\cite{wortwein2015multimodal, chollet2015exploring, chollet2016multimodal, wortwein2015automatic}. These systems provide feedback through responsive virtual audience behaviors, such as nodding or smiling of the virtual avatars when eye contact is made by the speaker~\cite{chollet2015exploring}, or through visual indicators, such as a progress bar that turns green to indicate sufficient eye contact~\cite{chollet2016multimodal}. However, these systems are designed for practice purpose in virtual environments and do not offer feedback during live, real-world presentations. Moreover, they primarily assess eye contact based on its overall duration relative to speech time. By contrast, our work focuses on real-time eye-contact assistance during live presentations. We not only provide in-situ feedback to the speaker but also go beyond duration-based metrics by evaluating the spatial distribution of gaze across audience members, thereby promoting more balanced and effective audience engagement.

\section{System Design}
In this section, we first give an overview of \name, and then detail the design of \name.

\subsection{Design Overview}


\subsubsection{Design Goal}

The goal of \name is to support speakers in maintaining focused and balanced eye-contact behavior during live presentations. To achieve this, we design \name as a wearable system that provides real-time, in-situ feedback based on continuous monitoring and analyzing of the speaker’s gaze behavior. The system is guided by two primary objectives: (1) ensuring the speaker maintains attention on the audience, and (2) promoting an even distribution of attention across audience members. Accordingly, \name is designed to detect and respond to the following ineffective eye-contact patterns:

\begin{itemize}[leftmargin=*, wide, labelwidth=!, labelindent=0pt]
\setlength\itemsep{0.3em}
    \item \textit{Insufficient eye contact.} When the speaker rarely looks at the audience, e.g., fewer than 20\% of the speaker's gazes are directed to the audience, \name prompts the speaker to engage visually with the audience.

    \item \textit{Imbalanced attention.}  When certain audience regions receive significantly less attention, \name advises the speaker to redirect gaze to the under-engaged audience to achieve balanced attention distribution across all audience members. 
\end{itemize}

\begin{figure}[t]
    \centering
    \includegraphics[scale=0.43]{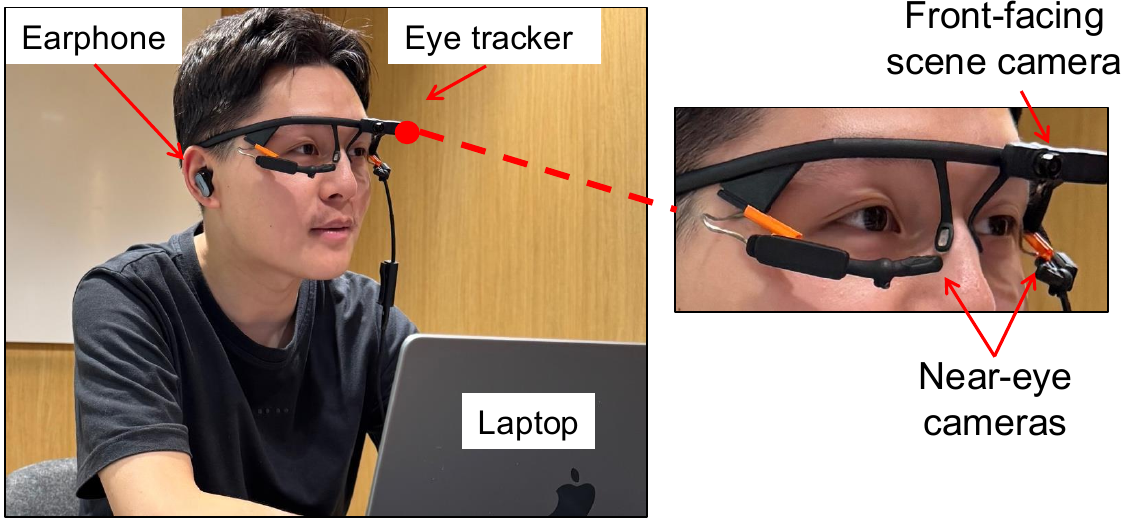}
    \caption{Hardware platform for \name. The speaker wears an eye tracker and an earphone, both connected to the laptop. The eye tracker has a front-facing scene camera and two near-eye cameras.
    }
    \label{fig:hardware}
    \vspace{-0.1in}
\end{figure}

\begin{figure*}[t]
	\centering
    \includegraphics[scale=0.53]{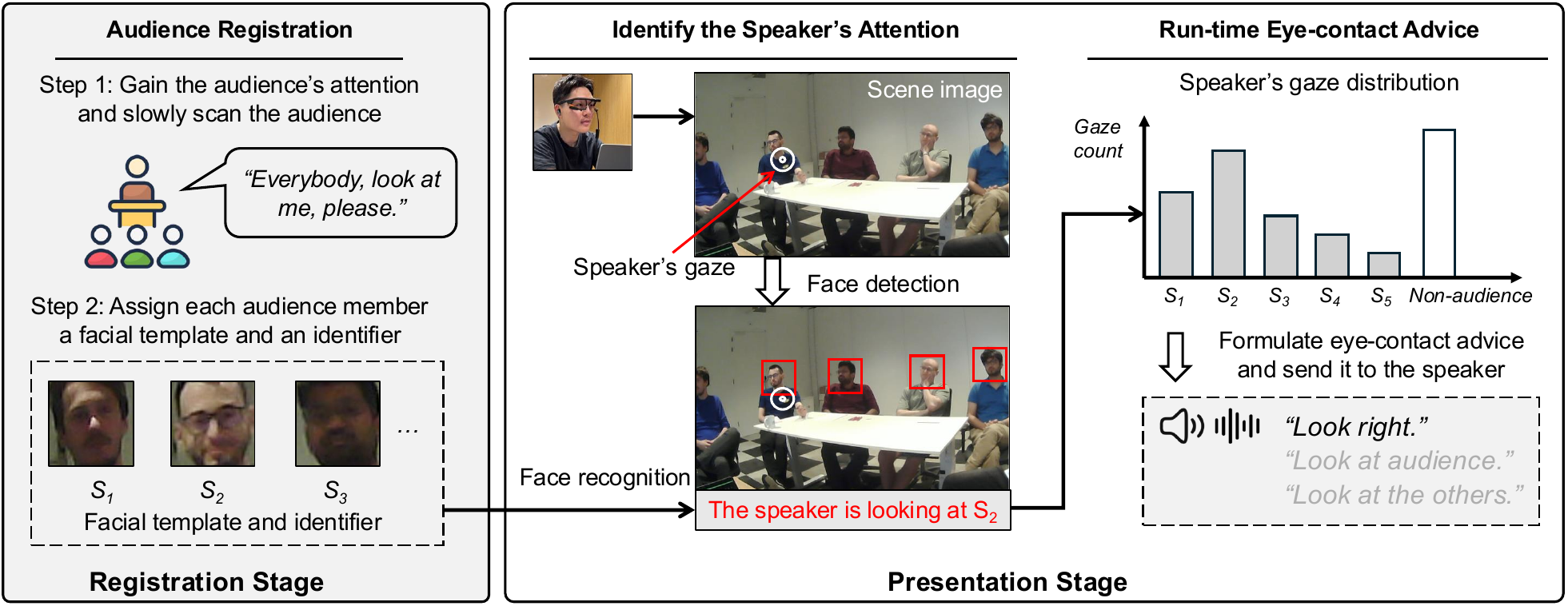}
    \caption{The workflow of \name. Before the presentation, \name performs audience registration. 
    Specifically, the speaker first captures the audience's attention, then uses the eye tracker's scene camera to scan and record all audience members. After that, \name detects faces in the video and assigns a facial template and a unique identifier to each audience member. In the presentation stage, \name continuously tracks the speaker's gaze distribution across the audience members and non-audience regions. 
    By analyzing this gaze distribution in real time, \name detects ineffective eye-contact patterns, generates appropriate eye-contact feedback, and discreetly delivers it to the speaker.}
    \label{fig:overview}
\end{figure*}

\subsubsection{Hardware Platform}
We implement \name using commercially available hardware: a wearable eye tracker, a laptop, and a single earphone. As shown in Figure~\ref{fig:hardware} (a), the speaker wears the eye tracker and an earphone, both connected to the laptop. The roles of each component are described below: 

\begin{itemize}[leftmargin=*, wide, labelwidth=!, labelindent=0pt]
\setlength\itemsep{0.3em}
    \item \textit{Eye tracker.} We use the Pupil Core eye tracker~\cite{kassner2014pupil} to capture the speaker's gaze and the scene view to track the speaker's attention. As shown in Figure~\ref{fig:hardware}, the eye tracker has one front-facing scene camera and two near-eye cameras. The scene camera captures scene view at a resolution of $1280\times720$, while the near-eye cameras capture the speaker’s eye images for gaze estimation at 120 Hz. Using the open-source Pupil Capture platform, we obtain the speaker’s gaze positions as pixel coordinates in the scene image.

    \item \textit{Laptop:} The \name software runs on the speaker’s laptop, which is typically used to display the presentation slides. In this study, we use a 2021 MacBook Pro with an M1 Pro chip, representing a standard general-purpose laptop rather than a high-end device. The software of \name analyzes the gaze signal and scene video captured by the eye tracker and generates real-time eye-contact advice based on the speaker’s gaze distribution.
    
    \item \textit{Earphone:} A single Bluetooth-connected earphone is used to deliver audio feedback to the speaker during the presentation. While our setup is simple and effective, alternative feedback channels, such as near-eye displays~\cite{hallidayGlasses,vuzixGlasses} in smart glasses or augmented reality devices, can also be used to replace the earphone.
\end{itemize}

\subsubsection{Workflow of \name}

Figure~\ref{fig:overview} illustrates the workflow of \name. The system operates in two main stages: audience registration before the presentation and gaze monitoring with feedback during the presentation.

\vspace{6pt}
\noindent
\textbf{Registration Stage}. Before the presentation begins, \name performs audience registration to assign each audience member a unique identifier and a facial template for later identification. To initiate this process, the speaker asks the audience to look toward them and then scans the room to record a short video capturing all audience members. This step is necessary because audience members are often spread out in the room, making it difficult to capture them all in a single frame. \name processes the scene video to detect individual faces and assigns each audience member a unique identifier, from $S_1$ to $S_N$, based on their spatial left-to-right position in the scene view, where $N$ is the total number of audience members. Each audience is also associated with a facial image, extracted from the video, to be used as a template for face identification during the presentation.


\vspace{6pt}
\noindent
\textbf{Presentation Stage}. During the presentation, \name continuously tracks the speaker’s gaze distribution, which reflects how the speaker allocates visual attention across different audience members and non-audience regions. For each incoming scene image and corresponding gaze point from the eye tracker, \name first performs face detection and determines whether the speaker is looking at an audience member by measuring the distance between the gaze point and the centers of detected faces. If the speaker is indeed looking at the audience, \name invokes a face identification module (Section~\ref{sec:FaceIdentification}) tailored for live speaking scenarios to determine which audience member is being looked at. The system updates the speaker’s gaze distribution in real time. Based on this distribution, \name generates eye-contact prompts to help the speaker maintain attention on the audience and promote balanced gaze allocation (Section~\ref{sec:EyeContactAdvice}). These eye-contact prompts are delivered discreetly to the speaker through the earphone.


\subsection{Robust and Efficient Face Identification in the Scene Video via Anchor Tracking}
\label{sec:FaceIdentification}

To generate accurate real-time eye-contact advice, \name requires an up-to-date gaze distribution that reflects how the speaker allocates visual attention across audience members and non-audience regions. This, in turn, depends on fast and reliable identification of the audience members present in each scene image. Specifically, this task involves detecting all faces in the scene view and identifying the one the speaker is currently looking at, referred to as the \textbf{\textit{target face}}. Formally, a detected face is considered the target face if the distance between the speaker’s gaze point and the center of the face is below a predefined threshold $L$. This is designed to compensate for minor gaze estimation errors of the eye tracker. If no target face is detected in the scene image, the speaker is considered to be looking at a non-audience region (e.g., the laptop). Below, we first outline the challenges in reliably identifying the target face under real-world presentation dynamics, followed by the details of our proposed solution.



\begin{figure*}[]
	\centering
    \includegraphics[scale=0.61]{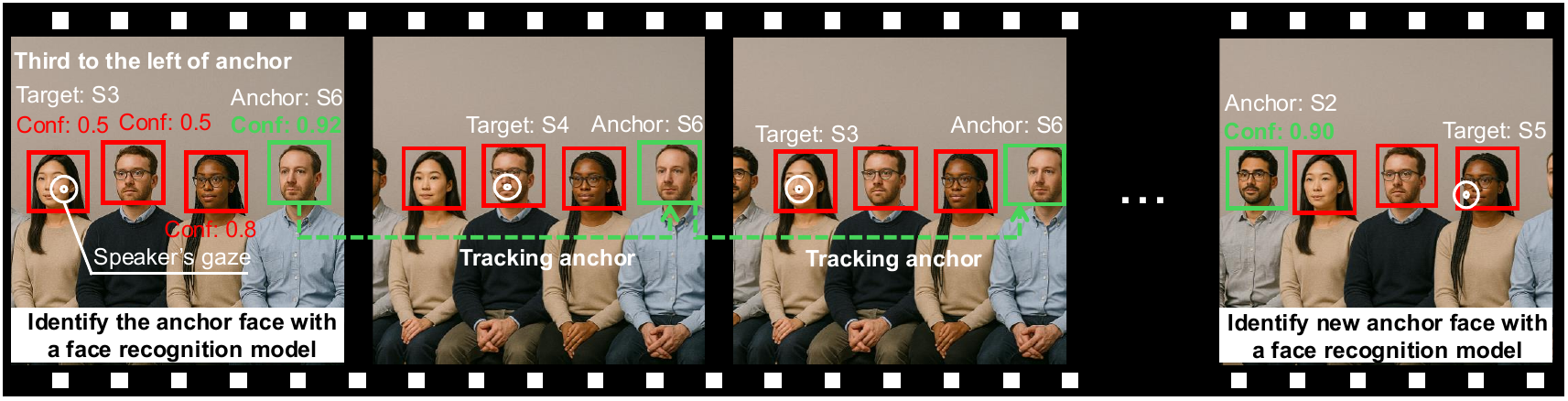}
    \caption{An illustration of the proposed face identification method in the scene video. We introduce the concept of \textbf{\textit{anchor face}}, which is a face that can be confidently identified either by the face identification model or by a face tracking algorithm. As an example, in the first frame, we apply the face identification model to all detected faces. The rightmost face, which receives the highest confidence score of $0.92$ (on a scale from $0$ to $1$), is selected as the anchor face (highlighted with a green bounding box). Once identified, the anchor face is tracked across subsequent frames (e.g., the second and third frames) until tracking fails --- for instance, when the anchor face disappears from the scene view in the last frame. Now, as long as the anchor face can be successfully tracked or identified, we can infer the identity of the target face by its relative spatial position to the anchor face. For example, in the first frame, the anchor face is identified as $S_6$. Since the target face lies three positions to the left of the anchor face, it is inferred as $S_3$. If tracking of the anchor face fails or it is no longer visible, we re-execute the face recognition model to select a new anchor face on the current scene frame and continue the process.}
    \label{fig:facerecog}
\end{figure*}

\subsubsection{Challenge}
\label{sec:challenge}

Accurately identifying the target face in the scene view with low latency on a standard laptop is a critical requirement for \name. A common approach involves first applying a pre-trained face detection model, such as YOLOv8n-face~\cite{yolov8}, followed by using a face identification model like Inception-ResNet-V1~\cite{inceptionresnet} to extract a facial representation of the detected face. This representation is then compared against the registered facial templates to determine the identity of the audience member. However, this conventional pipeline often results in poor identification accuracy due to substantial discrepancies between the audience’s facial appearances during the presentation and their corresponding templates captured during the registration stage. 

Specifically, these discrepancies stem from several real-world factors. First, the speaker frequently changes head pose while shifting visual attention among different audience members, the laptop, and the display screen, resulting in varying viewing angles of audience faces in the scene images. Second, audience members themselves often move or turn their heads during the presentation. Moreover, the audience's faces are sometimes partially occluded by hands. For example, when they rest their chin on their hands, adjust their glasses, scratch their heads, or place a hand on their foreheads. Together, these pose variations and occlusions introduce significant appearance differences 
that make direct face identification from scene images particularly challenging.

\subsubsection{Our solution}
To address this challenge, we introduce the concept called \textbf{\textit{anchor face}}, which refers to the face in the scene that can be confidently identified either by a face identification model or a face tracking algorithm. 

As illustrated in Figure~\ref{fig:facerecog}, in the first frame of the scene video, we apply the face identification model to all detected faces. The rightmost face $S_6$, marked by a green bounding box, is recognized with a high identification confidence score of $0.92$ (on a scale from $0$ to $1$) and is selected as the anchor face. We then \textit{track} the anchor face across subsequent frames (e.g., the second and third frames) using a position-based face tracking algorithm, \textit{which relies on the spatial continuity of detected faces rather than facial appearance.} This allows for robust face tracking even in the presence of appearance changes or partial occlusions. The anchor face continues to be tracked until it disappears from the scene view, as shown in the final frame of Figure~\ref{fig:facerecog}. At that point, a new anchor face is selected by re-executing the face identification model.


Once the anchor face is established in a given scene image, we can infer the identity of the target face based on its relative spatial position to the anchor. This leverages the observation that the position layout of the audience remains stable during the presentation. For instance, in the first frame of Figure~\ref{fig:facerecog}, the target face is the third face to the left of the anchor face, and therefore, we can infer its identity as $S_3$. These relative positions are determined during the registration stage (see Section~\ref{sec:detailDesignFaceRecog}). 

Moreover, by avoiding the computationally expensive face identification model on every scene image and instead relying on anchor face tracking and position-based inference, our approach significantly reduces system latency and enables higher frame processing rates. This results in slight shifts in face positions across consecutive scene images, which in turn enhances the robustness of the anchor face tracking. As we will demonstrate in later sections, our approach achieves robust face identification by effectively mitigating the impact of appearance variations through anchor face tracking and position-based identity inference. 

\subsubsection{Detailed Design} 
\label{sec:detailDesignFaceRecog}
We now detail the proposed method to identify the target face in the scene video of the speaker.

\begin{figure}[]
    \centering
    \includegraphics[scale=0.34]{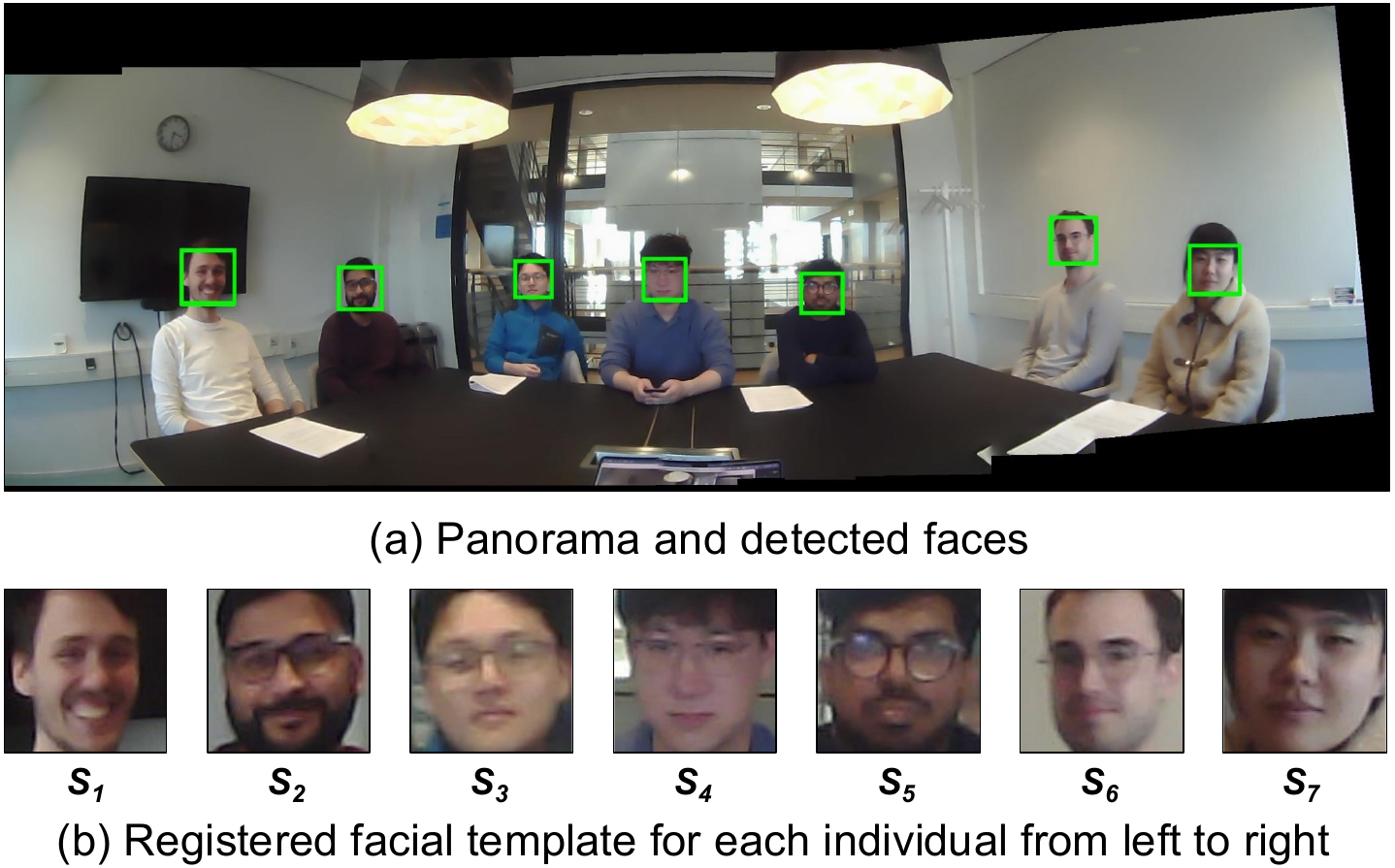}
    \caption{The panorama built from the scene images captured by the eye tracker and the facial templates with the unique identifiers registered for all the audience members by their relative positions from left to right.
    }
    \label{fig:Panorama}
    \vspace{-0.1in}
\end{figure}

    
\vspace{6pt}
\noindent
\textbf{{Inferring the relative positions of audience members.}} 
During the registration stage, \name records a scene video as the speaker slowly scans the audience from left to right. Then, \name samples $M$ frames from the video to construct a panorama, as illustrated in Figure~\ref{fig:Panorama} (a). After that, it applies the face detection model, a pre-trained YOLOv8n-face~\cite{yolov8}, to the panorama and determines the relative positions of audience members based on the horizontal locations of their detected faces. As shown in Figure~\ref{fig:Panorama} (b), each detected face is assigned a unique identifier: the leftmost face is labeled $S_1$, and the rightmost is labeled $S_{N}$, where $N$ is the total number of detected faces. The registered facial template with the identifier is illustrated in Figure~\ref{fig:Panorama} (b), which will be used by the face identification model in the presentation stage to determine the anchor face. Note that, to ensure reliable recognition during the presentation, \name does not use distorted faces from the stitched panorama as templates. Instead, it extracts the corresponding face from the original sampled frame (used in panorama construction) to serve as the facial template, as shown in Figure~\ref{fig:Panorama}(b).


\vspace{6pt}
\noindent    
\textbf{{Selecting the anchor face.}} The anchor face is a face in the current scene image that can be confidently identified using either a face identification model or a face tracking algorithm. If no anchor face was available in the previous frame, such as in the very first frame of the presentation, \name applies a face identification model (e.g., Inception-ResNet-V1~\cite{inceptionresnet}, pre-trained on the VGGFace2 dataset~\cite{cao2018vggface2}) to all detected faces in the current frame. It selects the face with the highest identification confidence as a candidate anchor. If the confidence score exceeds a threshold $p$, this candidate is accepted as the anchor face; otherwise, \name defers selection to the next frame. 
    
If an anchor face was available in the previous frame, as shown in the second and third frames of Figure~\ref{fig:facerecog}, \name attempts to track it in the current frame. It does so by selecting the face that is spatially closest to the previous anchor face as the candidate. If the distance between the two is less than a threshold $D$, the candidate is assigned as the anchor face; otherwise, e.g., when the anchor face disappears from the current frame (as in the last frame of Figure~\ref{fig:facerecog}), \name considers the anchor face lost and re-initiates the selection process using the face identification model. 

    

\vspace{6pt}
\noindent
\textbf{{Identifying the target face.}} Once an anchor face is selected, \name infers the identity of the target face in the same frame by computing its relative position with respect to the anchor face. For example, in the first frame of Figure~\ref{fig:facerecog}, the anchor face is identified as $S_6$ and the target face lies three positions to its left; the system infers the identity of the target face as $S_3$.

\subsubsection{Effectiveness}
Below, we evaluate the effectiveness of our face identification method in a real-world setting.

\textbf{Setup.} We recruit eight participants for the evaluation, including one speaker and seven audience members. The speaker wears the eye tracker and engages in a one-minute conversation with the audience while the scene view is recorded. Both the speaker and the audience can move their body naturally. 
As a baseline for comparison, we use the straightforward approach described in Section~\ref{sec:challenge}. Specifically, we apply the same face identification model used in our method, i.e., Inception-ResNet-V1~\cite{inceptionresnet}, to extract the feature representation of the target face and compare it against all registered facial templates. If the similarity between the representations of the target face and any template exceeds a threshold $D$, the target face is assigned the identity of this facial template. We consider \textit{face identification accuracy} and \textit{processing latency} as the evaluation metrics. The recorded video contains 855 scene images. We evaluate the face identification accuracy across \textit{all detected faces} in these scene images using manual annotations as ground truth. We manually annotate the identities of the detected faces in each scene image. For processing latency, we measure the time from receiving the scene image to producing the identification results.

\begin{figure}[]
	\centering
    \includegraphics[scale=0.42]{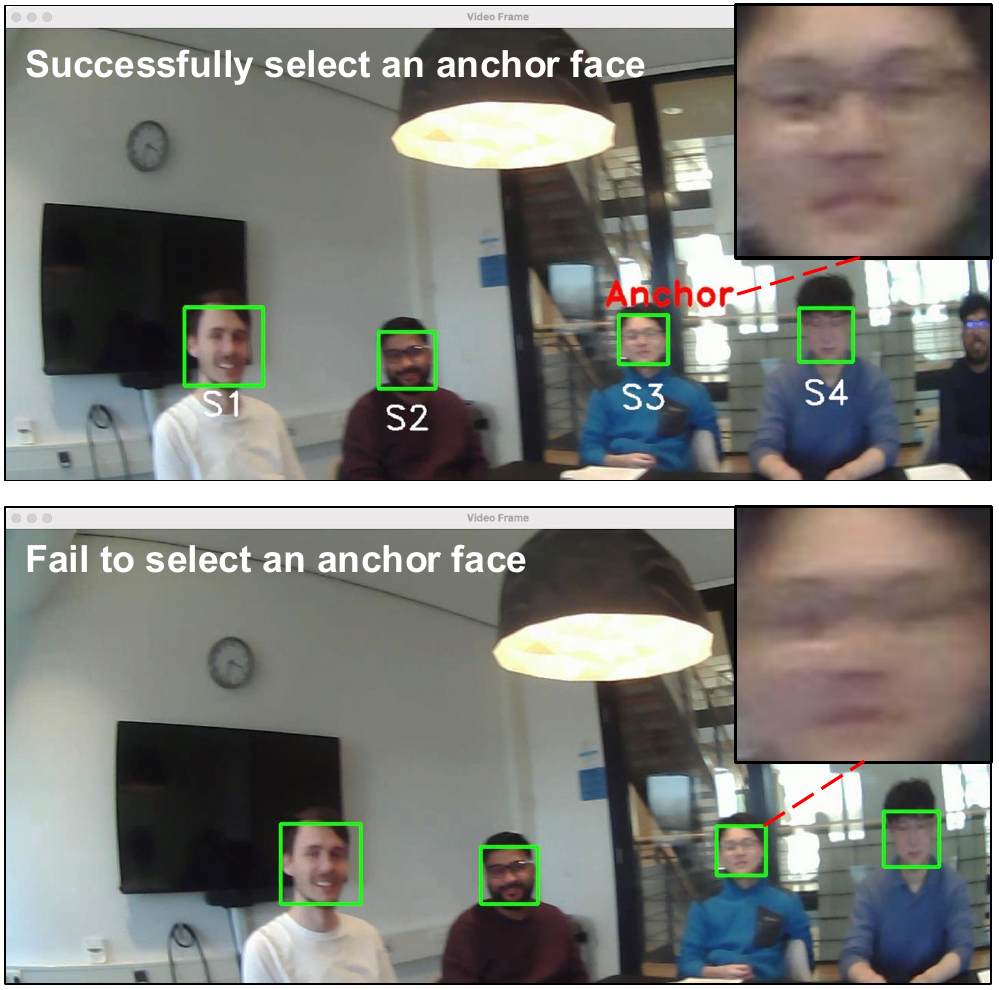}
    \caption{Illustration of face identification examples in real-world scenarios. The top image shows a frame where \name successfully selected the anchor face. In the bottom (consecutive) frame, \name fails to track the anchor face due to significant face position shifts and is unable to select a new anchor face because of motion blur, particularly illustrated in the top-right region.
    }
    \label{fig:failurecase}
    \vspace{-0.1in}
\end{figure}


\textbf{Results.} Table~\ref{Tab:effectiveness} reports the face identification accuracy and the average latency per frame for both the proposed method and the baseline. Our method achieves significantly higher accuracy with lower latency. A demo video showcasing the face identification performance of our proposed method is available at this Google Drive link\footnote{Visualization of face identification results: \url{http://bit.ly/4lzBums}}. Most misclassifications by the proposed method occur when the speaker moves their head rapidly, causing \name to lose track of the anchor face. An example is shown in Figure~\ref{fig:failurecase}, where the bottom image is the consecutive frame of the top one. In the top frame, \name successfully selects an anchor face. However, in the subsequent frame, it fails to track the anchor face due to significant shifts in face positions. Additionally, motion blur, particularly visible in the top-right region of the bottom frame, reduces the confidence of the face identification model, preventing \name from selecting a new anchor face. In such cases, \name treats these frames as unreliable and refrains from assigning any identity, which contributes to the observed identification errors. Notably, motion blur also causes the baseline method to fail in identifying the audience members.

\begin{table}[]\tiny
\caption{The face identification accuracy (in percentage) across all detected faces in the recorded scene video, and the average per-frame latency (in ms) for both the proposed and the baseline methods. Compared to the baseline, the proposed method achieves substantially higher accuracy while having a significantly lower latency in real-world dynamic speaker-audience interactions.}
\label{Tab:effectiveness}
\resizebox{0.925\linewidth}{!}{%
\begin{tabular}{c|cc}
Method            & Accuracy (\%) & Latency (in ms) \\ \hline
Proposed method   & 93.3     & 28        \\
Baseline method & 30.8     & 57       
\end{tabular}}
\vspace{-0.1in}
\end{table}

In terms of efficiency, \name achieves significantly lower processing latency than the baseline, as it applies the face identification model to only 6\% of the scene images. Specifically, for one scene image, the face identification model takes $30$ ms on the evaluation device (a 2021 MacBook Pro with an M1 Pro chip), whereas the face tracking algorithm requires only $0.1$ ms, enabling faster and more scalable processing.

\subsection{Generating Eye-contact Advice}
\label{sec:EyeContactAdvice}

Building on the proposed face identification method, \name computes the speaker’s gaze distribution across audience members and non-audience regions. This distribution forms the basis for detecting specific situations in which the speaker’s eye-contact behavior is ineffective and for generating appropriate feedback.

\subsubsection{Computing speaker's attention distribution} 
Consider a time window of $P$ seconds containing $X$ consecutive scene images and gaze points. Let $\bar{X}$ denote the number of scene images where a target face is selected, and let $X_i$ denote the number of scene images in which the target face is identified as $S_i$. We define \textbf{the eye-contact proportion} ($EP$) in this period as: \begin{equation}
    EP=\bar{X}/X\cdot100\%.
\end{equation}
This metric reflects the percentage of frames in which the speaker directs attention towards the audience. We also define \textbf{the eye-contact distribution} ($ED$) over a specific audience member $S_i$ in this period as: \begin{equation}
    ED(S_i)=X_i /\bar{X} \cdot100\%.
\end{equation}
$ED$ indicates the proportion of the speaker’s attention directed at $S_i$ relative to the total attention directed at all the audience members. 

Based on these metrics, \name detects two specific ineffective eye-contact situations and generates corresponding feedback to support the speaker.

\subsubsection{Eye-contact Situations and Advice} 

In the current design, \name detects and responds to two specific types of eye-contact patterns during a presentation:

\begin{itemize}[leftmargin=*, wide, labelwidth=!, labelindent=0pt]
\setlength\itemsep{0.3em}
    
    \item \textit{Insufficient eye contact:} refers to situations where the speaker rarely looks at the audience. \name checks for this condition every $n$ seconds. At the end of each detection period, if the $EP$ is below a threshold $r_p$, the system determines that the speaker is not engaging the audience enough and lacks eye contact. \name will generate the advice, ``\textit{look at the audience}''.

    \item \textit{Imbalanced attention:} represents the situation where the speaker pays significantly less attention to certain audience. \name checks this situation every $k$ seconds and identifies the audience member with the lowest eye-contact distribution. To provide simple and actionable feedback to the speaker, \name generates a simple advice to encourage the speaker to look at the less-engaged audience. Specifically, if this individual is on the left side of the audience, \name advises: ``\textit{look left more}''; otherwise, it advises: ``\textit{look right more}''.

\end{itemize}
We empirically set $r_p=20\%$, $n=30$ seconds, and $k=75$ seconds. The generated advice is delivered to the speaker via an earphone connected to the laptop through Bluetooth. We convert the textual advice into the audio prompt using the pyttsx3~\cite{pyttsx} text-to-speech conversion library in Python.

\subsection{User Interface Design}
\begin{figure}[]
	\centering
    \includegraphics[scale=0.34]{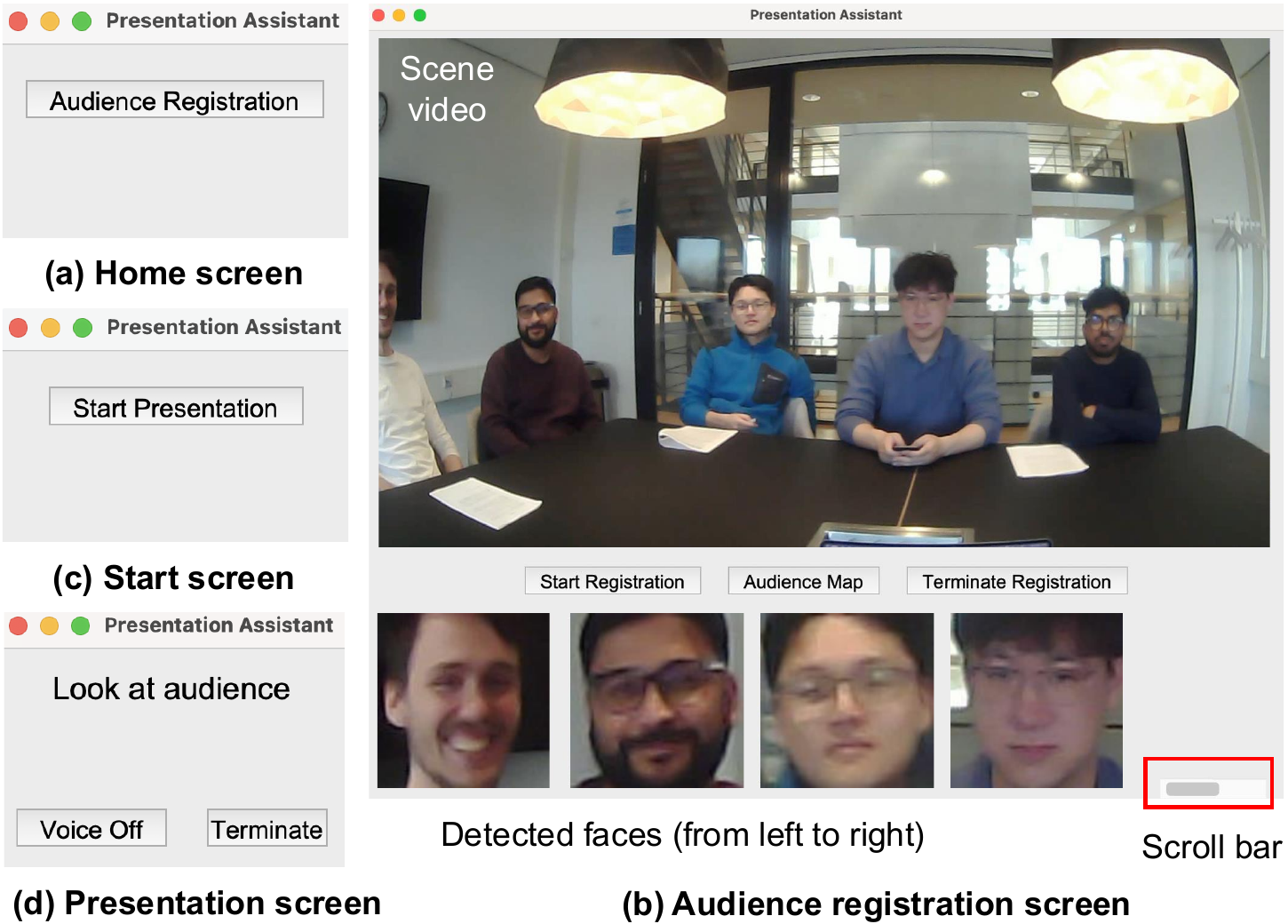}
    \caption{The User Interface (UI) for \name. The UI contains four screens: (a) Home screen; (b) Audience registration screen; (c) Start screen; and (d) Presentation screen.
    }
    \label{fig:UIDesign}
    \vspace{-0.1in}
\end{figure}

We design the user interface (UI) for \name using Tkinter~\cite{tkinter}. As shown in Figure~\ref{fig:UIDesign}, the UI includes four UI screens. We detail each of them below. 

\begin{itemize}[leftmargin=*, wide, labelwidth=!, labelindent=0pt]
\setlength\itemsep{0.3em}
    
    \item \textit{Home screen.} After launching the system, the speaker is presented with the home screen (Figure~\ref{fig:UIDesign}(a)). Tapping the ``Audience Registration'' button will navigate to the audience registration screen.

    \item \textit{Audience registration screen}. This screen is used during the audience registration stage. As shown in Figure~\ref{fig:UIDesign}(b), the scene video appears at the top, with three control buttons in the middle. The speaker begins by clicking ``Start Registration'', which then toggles to ``Stop Registration''. While the recording is active, the speaker slowly scans the audience to capture all members. Once complete, the speaker clicks ``Stop Registration'' to end the recording. Next, the speaker clicks the ``Audience Map'' button to register each audience member with a template facial image and a unique identification number. These template facial images are displayed at the bottom of the screen, arranged from left to right based on the audience members’ relative positions. A scroll bar is available to navigate through the list when the number of audience members exceeds the visible area. Finally, the speaker clicks ``Terminate Registration'' to proceed to the ``Start screen''. 

    \item \textit{Start screen}. This transitional screen appears after registration and before the presentation begins. It features a compact layout, enabling the speaker to open presentation slides and reposition the interface as needed. When ready, the speaker clicks ``Start Presentation" to begin.

    \item \textit{Presentation screen.} This screen is active during the presentation. The generated eye-contact advice is displayed at the top of the screen. Two buttons are located at the bottom: the ``Voice Off'' button mutes the audio advice, and the ``Terminate'' button is used to exit \name.

\end{itemize}

\section{Evaluation}

In this section, we evaluate the performance of \name by conducting a user study with both speakers and audience members in real-world presentation scenarios.


    


\subsection{Evaluation Setup}

\begin{table}[]
\caption{Demographics information of the 28 participants.}
\label{Tab:demogra}
\begin{tabular}{ll}
\Xhline{2\arrayrulewidth}\\[-2.5ex]
Total number & 28 (4 speakers; 24 audience members)                                                                                       \\
Gender       & Female (4), Male (22)                                                                     \\
Age          & Min: 20, Max: 32                                                                          \\
Ethnicity    &East Asian (18), European (7), South Asian (3)\\ \Xhline{2\arrayrulewidth}
\end{tabular}%
\vspace{-0.1in}
\end{table}

\subsubsection{Participants} We recruit 28 participants for the study. Their demographics are shown in Table~\ref{Tab:demogra}. We divide the participants into four groups, each comprising one speaker and six audience members. Each speaker is instructed to deliver an approximately eight-minute presentation using self-selected slides that they are familiar with. The study is approved by the ethical committee of our institution. 

\subsubsection{User Study Design}
\begin{figure}[t]
	\centering
    \includegraphics[scale=0.3]{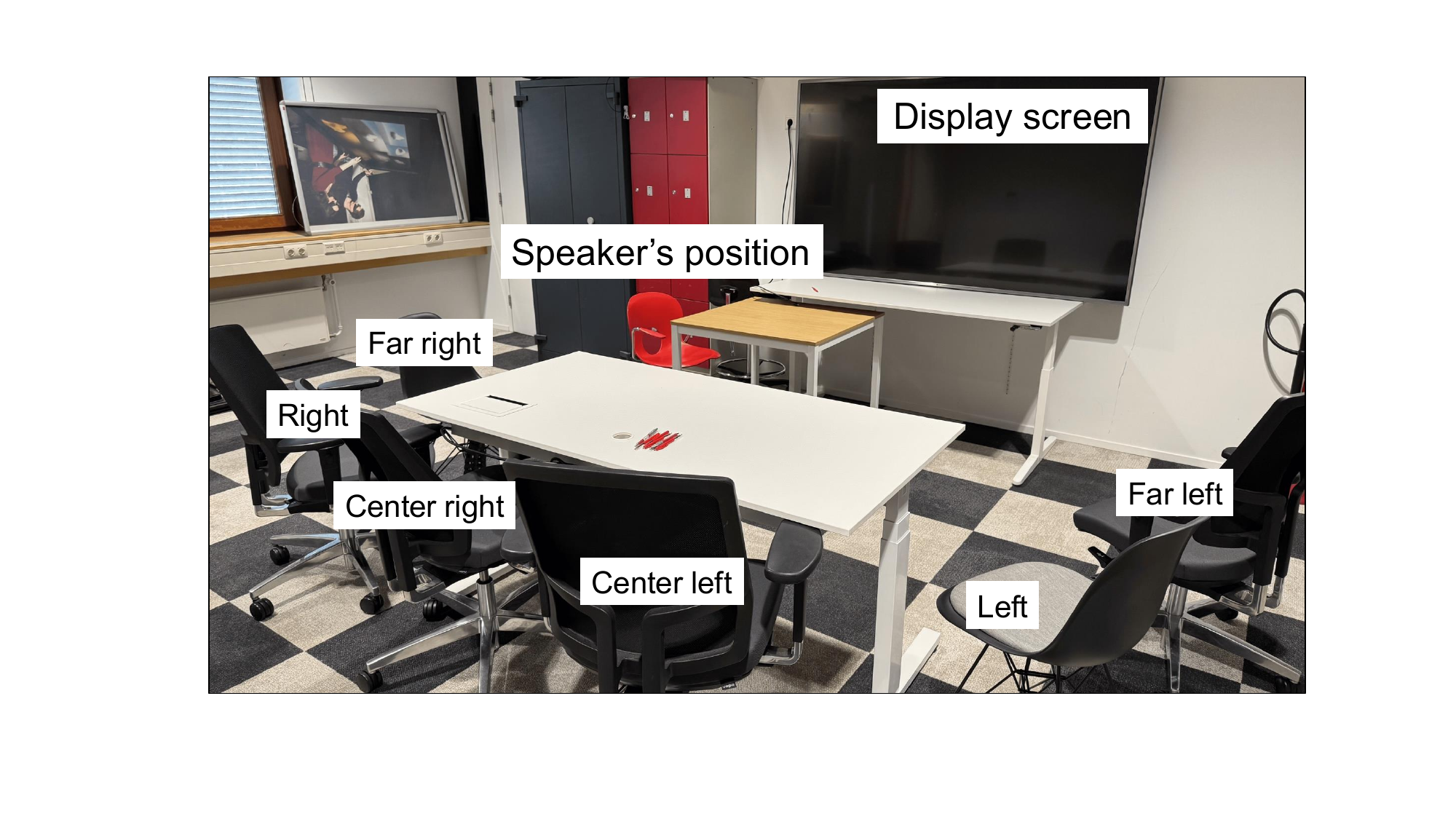}
    \caption{The meeting room for the user study. The speaker is free to stand or sit in front of the display screen to give the presentation, while the audience take the chairs arranged from far left to far right (from the speaker's view). The display screen is used to show the presentation slides. }
    \label{fig:UserStudyRoom}
    \vspace{-0.1in}
\end{figure}

The user study is conducted in a meeting room, as shown in Figure~\ref{fig:UserStudyRoom}. In the presentation, the speaker is free to stand or sit in front of the display screen, while the audience occupies the black chairs arranged from left to right from the speaker's perspective. The laptop is connected to a display screen to show the presentation slides. Each group participates in two sessions, with the presentation divided into two four-minute parts, one delivered per session. In both sessions, the speaker wears an eye tracker, and both gaze data and scene video are captured. 
Prior to each session, the eye tracker is calibrated for the speaker using the calibration software provided by the eye tracking vendor. 

In the first session, the speaker delivers the first part of the presentation without using \name. In the second session, the speaker delivers the second part with the assistance of \name. This setup allows us to examine whether \name improves the speaker’s eye contact with the audience by analyzing and comparing both the objective data recorded from the two sessions, as well as the subjective feedback from all participants. To ensure the effectiveness of the subjective feedback, 
the audience is instructed to pay close attention to any changes in the speaker’s eye contact between the two sessions.

To avoid influencing the speaker’s natural eye-contact behavior, we do not disclose the specific purpose of the study to the speakers prior to the first session. Instead, we ask the speakers to wear the eye tracker and deliver the presentation as usual, and inform them that their gaze behavior will be monitored and analyzed. Before the second session, we explain the full purpose of the study to the speakers and provide instructions on how to use \name. The speakers then deliver the second part of the presentation. 

After the second session, both the speaker and the audience are asked to complete a questionnaire (detailed in Section~\ref{Sec:evaluationmethod}) and are invited to provide open-ended feedback. The study for each group lasts approximately 30 minutes. 


\subsubsection{Evaluation Methodology}
\label{Sec:evaluationmethod}

We would like to explore the following research questions (RQs):

\begin{itemize}[leftmargin=*, wide, labelwidth=!, labelindent=0pt]
\setlength\itemsep{0.3em}
    \item \textbf{RQ1:} \textit{Does \name increase the speaker’s eye contact duration with the audience during the presentation?}
    
    \item \textbf{RQ2:} \textit{Does \name help the speaker distribute attention more evenly across audience members?}
    \item \textbf{RQ3:} \textit{Do speakers perceive \name as effective in improving their eye contact with the audience during presentations?}
    \item \textbf{RQ4:} \textit{Does the use of \name enhance audience engagement and make the presentation more interactive? 
    }
\end{itemize}

We introduce different performance metrics for the evaluation. Specifically, to address \textbf{RQ1}, we use \textit{eye contact proportion (EP)} as a quantitative indicator. The EP during a session is defined as the percentage of video frames (captured by the scene camera) where the speaker is looking at the audience. This metric serves as an approximation of the proportion of time the speaker makes eye contact with the audience throughout the session. A larger EP suggests that the speaker devotes more visual attention to the audience, and vice versa.

\begin{table}[]\huge
\caption{Questionnaire for speakers. Each question is answered based on a 5-point Likert scale.} 
\label{Tab:SpeakerSurvey}
\centering
\resizebox{0.93\linewidth}{!}{%
\begin{tabular}{lc}
\Xhline{3\arrayrulewidth}\\[-2ex]
\textbf{Question} & \textbf{Discription}\\
\\[-2ex]\Xhline{3\arrayrulewidth}\\[-2ex]
1. \textbf{Learnability}  &\makecell[l]{It is easy to use this system.}                                                                           \\ \\[-2ex]
2. \textbf{Attentiveness} &\makecell[l]{The speaker assistant helps me make \\more eye contact with the audience.}                                                                    \\ \\[-2ex]
3. \textbf{Balance} &\makecell[l]{The speaker assistant helps me balance \\my attention across the audience.}                                                             \\ \\[-2ex]
4. \textbf{Precision} &\makecell[l]{The speaker assistant correctly reflects my \\eye-contact situation, e.g., remind me to look\\ at the audience when I stare at my laptop.} \\ \\[-2ex]

5. \textbf{Usefulness}&\makecell[l]{The advice is overall useful for improving\\ eye contact with the audience.}                                                                   \\ \\[-2ex]
6. \textbf{Comfort} &\makecell[l]{The hardware eye-tracker does not affect \\my presentation process.}                                                                     \\ \\[-2ex]
7. \textbf{Intrusiveness} &\makecell[l]{The advice is non-intrusive during the \\presentation.}                                                                                                 \\ \\[-2ex]
8. \textbf{Willingness} &\makecell[l]{I am inclined to reuse this system for\\ improving eye contact in the future.}                                            \\ \\[-2ex]\Xhline{3\arrayrulewidth}
\end{tabular}%
}
\vspace{-0.1in}
\end{table}

For \textbf{RQ2}, we introduce a metric termed as \textit{gaze distribution entropy} (GDE), which quantifies the dispersion of the speaker’s gaze distribution across audience members during a presentation session. Let $X_i$ denote the number of frames in which the speaker is looking at the audience member $S_i$, and let $N$ be the total number of audience members. We define $X_{sum}=\sum_{i=1}^{N}X_i$. Then, the gaze distribution entropy is calculated as: \begin{equation}
    GDE = -\sum_{i=1}^{N}{\frac{X_i}{X_{sum}}\log \frac{X_i}{X_{sum}}},
\end{equation}where $X_i / X_{sum}$ represents the proportion of the speaker’s gazes directed at audience member $S_i$. A higher \textit{GDE} indicates a more evenly distributed attention across all audience members. For $N=6$, GDE ranges from $0$ to $1.79$. A $GDE$ of 0 indicates that the speaker focuses exclusively on a single individual, while a $GDE$ of 1.79 implies that the speaker distributes an equal amount of gazes among the audience members.

For \textbf{RQ3} and \textbf{RQ4}, we collected subjective feedback through post-study questionnaires completed by participants in each user study group. Speakers complete the questionnaire used in Table~\ref{Tab:SpeakerSurvey}. The first question evaluates the learnability of \name, followed by four questions assessing its usefulness in improving eye contact. The following two questions address potential negative effects during the presentation, and the final question asks about the speaker’s willingness to use \name in the future. Each question is answered and rated based on a 5-point Likert scale (1: strongly disagree; 2: disagree; 3: neutral; 4: agree; 5: strongly agree). {We perform the one-tailed Wilcoxon signed-rank test on the collected 5-point Likert scores.} 

Audience members complete a separate questionnaire shown in Table~\ref{Tab:AudienceSurvey}. The audience questionnaire begins by asking about the audience’s opinions on the \textbf{importance} of eye contact during presentations, followed by their observations of \textbf{differences} in the speaker’s eye-contact patterns between the two sessions. {The response to these two questions is either \textit{Yes} or \textit{No}. The next two questions ask which of the two sessions makes them feel more \textbf{engaged} and \textbf{interactive}. The audience is asked to make a binary choice between the two sessions, with the option to select \textit{Same}. They are also invited to provide open-ended comments.}

\begin{table}[t]
\caption{Questionnaire for the audience.}
\label{Tab:AudienceSurvey}
\resizebox{\linewidth}{!}{%
\begin{tabular}{lc}
\Xhline{3\arrayrulewidth}\\[-2ex]
\textbf{Question} & \textbf{Discription}\\
\\[-2ex]\Xhline{3\arrayrulewidth}\\[-2ex]
1. \textbf{Importance}  &\makecell[l]{Do you think eye contact between the speaker and \\the audience is important during a presentation?}                                                                           \\ \\[-2ex]
2. \textbf{Difference} &\makecell[l]{Do you think the speaker exhibited different eye \\contact patterns between the two sessions?}                                                                    \\ \\[-2ex]
3. \textbf{Engagement} &\makecell[l]{In which session did the speaker gain more of your \\ attention by making eye contact?}                                                             \\ \\[-2ex]
4. \textbf{Interaction} &\makecell[l]{In which session did the speaker make you\\ feel more interactive by making eye contact?}                                           \\ \\[-2ex]\Xhline{3\arrayrulewidth}
\end{tabular}%
}
\vspace{-0.1in}
\end{table}

\subsection{Results and Analysis}
Below, we discuss the evaluation results and findings for each of the research questions.

\vspace{3pt}
\noindent\textbf{Findings for RQ1: \name significantly increases the speakers' eye contact duration.}
\vspace{3pt}

\begin{figure}[]
	\centering
    \includegraphics[scale=0.45]{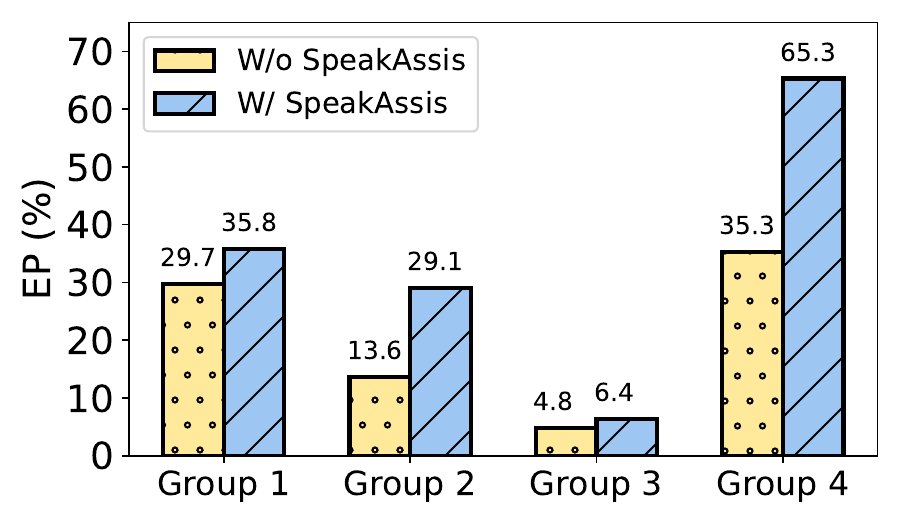}
    \caption{The eye contact proportion (EP, in percentage) of the speakers, with and without using \name. Overall, \name improves speakers’ eye contact across all four groups. On average, it achieves a 62.5\% increase in EP.
    }
    \label{fig:EF}
    \vspace{-0.1in}
\end{figure}

Figure~\ref{fig:EF} shows the eye contact proportion for the two sessions across the four groups. As mentioned, a larger value of EP suggests that the speaker pays more attention to the audience. In the first session, speakers give their presentations without using \name, while in the second session, they receive real-time eye-contact guidance from \name. 

Overall, for all four groups, the EP has increased when using \name, suggesting that \name effectively helps the speakers to make more eye contact with the audience. On average, it achieves a 62.5\% increase in eye contact proportion. Notably, the improvements in Groups 2 and 4 are more significant than those in Groups 1 and 3. We hypothesize that these differences may stem from variations in presentation content or speaker traits. For instance, after receiving feedback from \name, the speaker in Group 2 is more willing and able to sustain attention on the audience compared to the speaker in Group 3. Additionally, during the presentation, the speaker in Group 3 frequently needed to refer to and explain content on the slides, which reduced opportunities for maintaining eye contact with the audience. 

Moreover, the speaker in Group 2 remarked that ``\textit{One of my personal habits during presentations is that I sometimes face the audience without actually looking at them. \name can accurately detect this and remind me to look at the audience.}'' This feedback further supports the ability of \name in increasing the speaker’s eye-contact duration during presentations.

\vspace{3pt}
\noindent\textbf{Findings for RQ2: \name helps the speakers to distribute attention more evenly across audience members.} 
\vspace{3pt}

\begin{figure}[]
	\centering
    \includegraphics[scale=0.45]{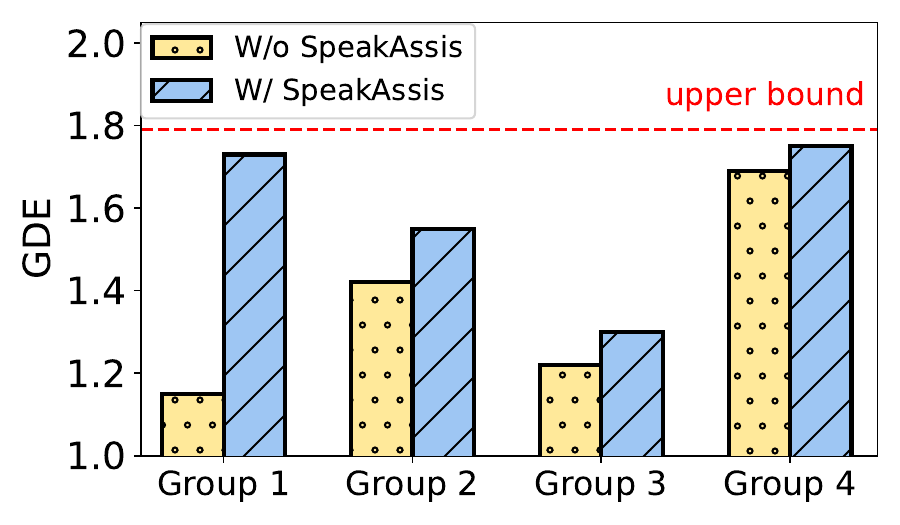}
    \caption{The gaze distribution entropy (GDE) of the speakers with and without using \name. Overall, \name encourages the speakers to distribute gazes evenly. On average, it achieves a 17.4\% increase in GDE.
    }
    \label{fig:ADE}
    \vspace{-0.1in}
\end{figure}

The {gaze distribution entropy} across the two sessions and four groups is shown in Figure~\ref{fig:ADE}. A higher \textit{GDE} indicates a more evenly distributed gaze across the audience, and vice versa. Overall, \name improves GDE in all four groups, which indicates that all speakers have a more balanced attention distribution when using \name. On average, it achieves a 17.4\% increase in GDE.

\begin{figure}[]
	\centering
	\subfigure[The speaker's gaze distribution among the audience in Group 1.]{\includegraphics[scale=0.43]{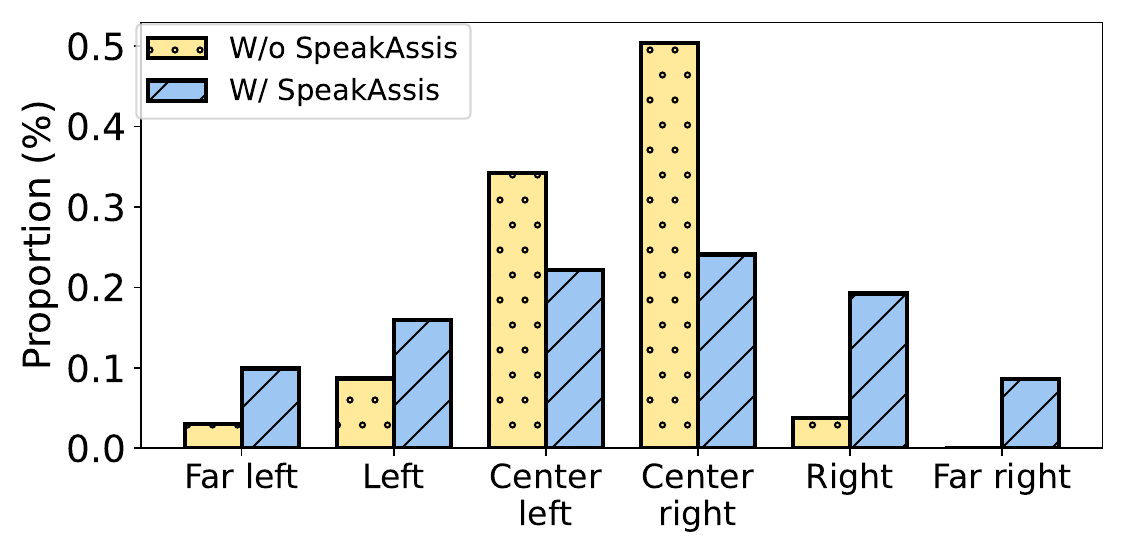}\label{Fig:G1Pro}} 
 \subfigure[The speaker's gaze distribution among the audience in Group 4.]{\includegraphics[scale=0.43]{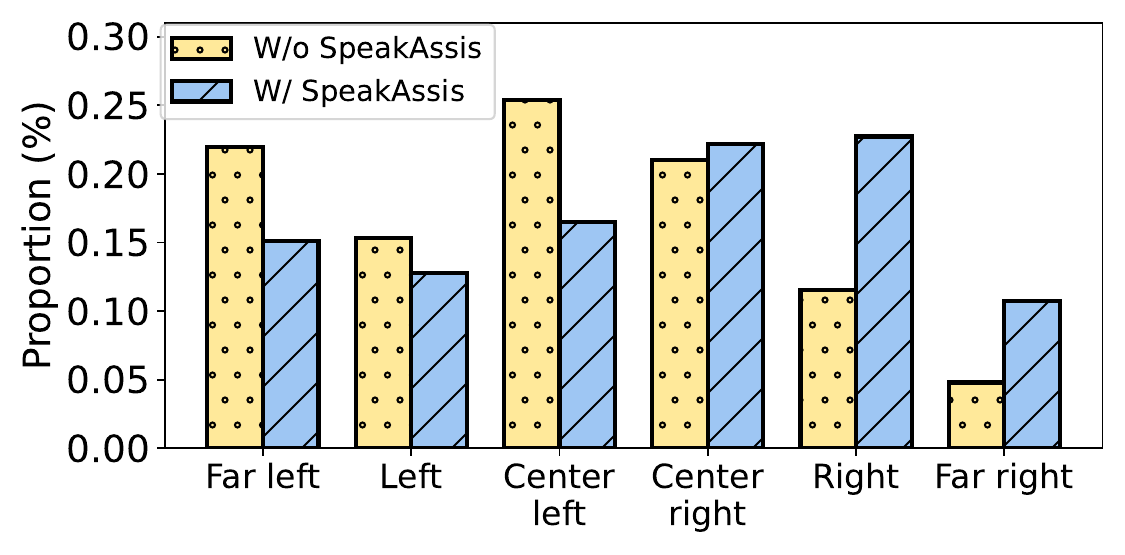}\label{Fig:G4Pro}}
\caption{The proportion of the speaker’s gaze (in percentage) directs toward audience members sitting in different positions in (a) Group 1 and (b) Group 4. In both groups, the speakers initially (w/o \name) direct less attention to individuals sitting on the right-hand side, i.e., ``Right'' and ``Far right'' positions. With the assistance of \name, their gaze distribution became more balanced, with attention more evenly directed across the audience.
}
\label{Fig:proportion}
\end{figure}

To gain deeper insight, Figure~\ref{Fig:proportion} shows the speakers' gaze distribution across the audience seated in different positions for Groups 1 and 4. As shown in Figure~\ref{Fig:G1Pro}, without using \name, the speaker in Group 1 tends to overlook audience members sitting on the right-hand side, specifically those in the ``Right'' and ``Far Right'' positions. After using \name and receiving the eye-contact guidance ``\textit{look right more},'' the speaker directs more attention toward those audience members, resulting in a much more balanced attention distribution across the audience.

Figure~\ref{Fig:G4Pro} shows that in Group 4, the speaker does not overlook any audience members during the first session (w/o \name), resulting in a higher gaze distribution entropy compared to other speakers (shown in Figure~\ref{fig:ADE}). However, similar to the speaker in Group 1, the speaker also allocates less attention to the audience sitting on the right-hand side without using \name. By contrast, after using \name in session two, the speaker distributes gaze more evenly across all audience members, bringing the gaze distribution entropy close to its maximum. This quantitative observation is further supported by the remark from an audience member: ``\textit{As a member in the corner, I got more attention in the second session},'' highlighting the effectiveness of \name in balancing the speaker's visual attention.

\vspace{3pt}
\noindent\textbf{Findings for RQ3: The speakers perceive \name as effective in improving their eye contact during presentations.}

\begin{figure*}[t]
	\centering
    \includegraphics[scale=0.46]{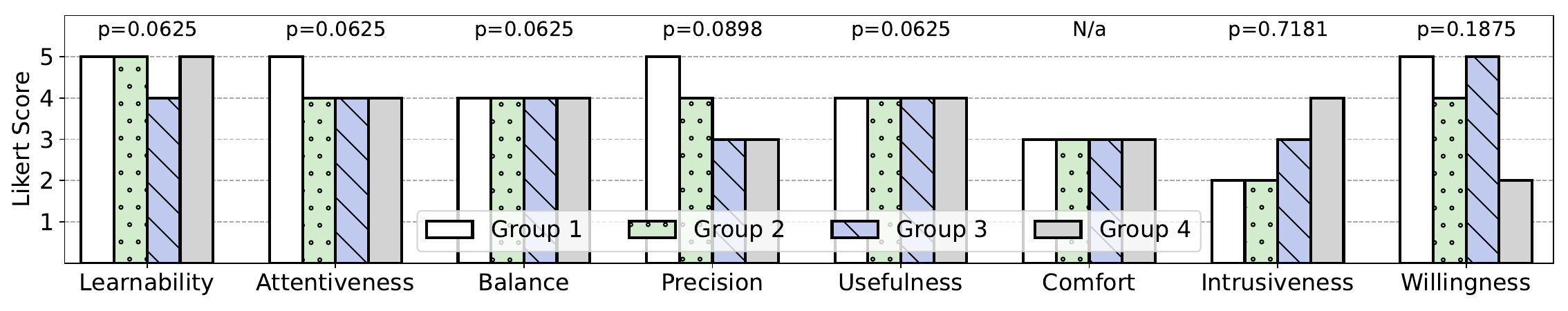}
    \caption{The response of speakers in different groups to the questionnaire in the user study and the $p$-value for one-tailed Wilcoxon signed-rank test for each question.
    }
    \label{fig:UserStudySpeakerAnswer}
\end{figure*}

We present the speakers' responses to the questionnaire in Figure~\ref{fig:UserStudySpeakerAnswer}. Overall, the results indicate that speakers generally agree on the following benefits of \name: (1) the system is easy to use (\textbf{Learnability}, $p=0.0625$); and (2) \name is helpful for improving eye contact (\textbf{Usefulness}, $p=0.0625$), particularly by increasing the duration of eye contact (\textbf{Attentiveness}, $p=0.0625$) and promoting more balanced attention across the audience (\textbf{Balance}, $p=0.0625$). This subjective feedback aligns with the quantitative results, which show improvements in both the proportion of eye contact (Figure~\ref{fig:EF}) and gaze distribution entropy (Figure~\ref{fig:ADE}) across all speakers. However, speakers also identify a limitation of the current implementation: the audio advice is perceived as intrusive during the presentation (\textbf{Intrusiveness}, $p=0.7181$). We acknowledge this as a limitation of the current prototype and discuss it further in Section~\ref{sec:discussion}. Nevertheless, three speakers expressed interest in using \name in the future (\textbf{Willingness}, $p=0.1875$). The speaker in Group 4, emphasize its strong potential as a training tool for novice speakers aiming to improve their eye contact skills.

\vspace{3pt}
\noindent\textbf{Findings for RQ4: \name enhances the audience engagement and the interactivity of the presentation.} 
\vspace{3pt}

Table~\ref{Tab:AudienceResponse} summarizes the audience's responses to the questionnaire in Table~\ref{Tab:AudienceSurvey}. Specifically, for the questions regarding \textbf{Importance} and \textbf{Difference}, the answer can be either \textit{Yes} or \textit{No}. For questions regarding \textbf{Engagement} and \textbf{Interaction}, the answer could be \textit{First Session}, \textit{Second Session}, or \textit{Same}. Table~\ref{Tab:AudienceResponse} summerizes the responses from all 24 audience members from the four groups. Overall, all audience members acknowledge the importance of maintaining eye contact during presentations. 



Moreover, as shown in Table~\ref{Tab:AudienceResponse}, among the 19 audience members who expressed a preference, 18 favored the presentation with \name, while only one preferred the presentation without it. To evaluate whether this preference is statistically significant, we perform a one-tailed binomial test~\cite{cox2018analysis}. The null hypothesis assumes that audience members have no preference between the two conditions, while the alternative hypothesis posits that audience members are more likely to prefer the presentation with \name. Using a significance level of \(\alpha = 0.05\), the test yields a \(p\)-value of $0.000038$, indicating that the observed result is highly unlikely under the null hypothesis. We therefore reject the null hypothesis and conclude that audience members perceive the presentation with \name as significantly more engaging and interactive.


To gain deeper insights, we conduct a follow-up interview with the audience member who reports feeling more engaged during the first session than the second. The interview reveals that this audience has limited interest in the presentation topic and quickly loses focus during the first session. This leads to a general sense of disengagement in both sessions, despite perceiving the first session as slightly more engaging. Notably, all audience members who report no perceived difference in \textbf{Engagement} or \textbf{Interaction} between the two sessions are from Group 4. The speaker in this group has more presentation experience than the other speakers. This is further supported by the results shown in Figures~\ref{fig:EF} and~\ref{fig:ADE}, where the speaker in Group 4 consistently achieves higher values in both EF and ADE across the two sessions, indicating more refined presentation skills and effective eye contact behavior.

\begin{table}[]
\caption{The summary of the response from the audience on the questionnaire. 
75\% of the audience feel more engaged and perceive the presentation as more interactive in the second session, where the speakers receive real-time eye-contact advice from \name.}
\label{Tab:AudienceResponse}
\begin{tabular}{cccc}
\Xhline{3\arrayrulewidth}\\[-2ex]
\textbf{Question} & \multicolumn{2}{c}{\textbf{Response}} &      \\
            & \textit{Yes}           & \textit{No}           &      \\ \\[-2ex]\hline\\[-2ex]
\textbf{Importance}  & 24            & 0            &      \\\\[-2.5ex]
\textbf{Difference}  & 19            & 5            &      \\ 
\\[-2ex]\hline\\[-2ex]
            & \textit{First Session}      & \textit{Second Session}     & \textit{Same} \\\\[-2.5ex]
\textbf{Engagement}  & 1            & 18            & 5    \\\\[-2.5ex]
\textbf{Interaction} & 1            & 18            & 5    \\ \\[-2ex]\Xhline{3\arrayrulewidth}
\end{tabular}

\end{table}

\subsection{Limitation and Future Research Direction}
 \label{sec:discussion}



A limitation of the current design is that the audio advice can be somewhat intrusive during live presentations. Several speakers noted that the audio cues, while helpful, occasionally disrupted their concentration, particularly when presenting complex or technical content. This highlights the need for alternative feedback modalities that are less disruptive yet still effective in guiding speaker behavior. A promising direction for future research is to explore non-intrusive feedback channels, such as near-eye displays that present subtle textual advice within the speaker’s peripheral vision. Examples such as the Halliday Glasses~\cite{hallidayGlasses} and the Vuzix Smart Glasses~\cite{vuzixGlasses}. Such displays could allow speakers to receive feedback without interrupting their verbal delivery or visual engagement with the audience. By leveraging more advanced wearable prototypes, we can make \name a more seamless, adaptive, and user-friendly system for enhancing public speaking through real-time feedback.

\section{Conclusion}
In this paper, we present \name, the first wearable system that provides real-time, in-situ advice to help speakers maintain effective eye contact during live presentations. With a head-mounted eye tracker, \name tracks and analyzes the speaker's gaze distribution among different audience members and the non-audience regions. Based on this, \name detects ineffective eye contact and generates appropriate advice to help the speaker maintain focused and balanced attention on the audience. We evaluate \name through a user study involving four speakers and 24 audience members. The quantitative results show that \name significantly improves the speaker's eye contact regarding eye contact duration and attention distribution. Furthermore, our statistical analysis from the audience's subjective feedback demonstrates that \name significantly enhances their engagement.


\balance

\bibliographystyle{ACM-Reference-Format}
\bibliography{09reference}


\end{document}